\documentclass[aps,prl,twocolumn,showpacs,superscriptaddress,groupedaddress]{revtex4}  % for review and submission
\usepackage{graphicx}  % needed for figures
\usepackage{dcolumn}   % needed for some tables
\usepackage{bm}        % for math
\usepackage{amssymb}   % for math

\begin{document}
% the following line is for submission, including submission to the arXiv!!
%\hspace{5.2in} \mbox{Fermilab-Pub-04/xxx-E}

\title{Slippery but tough - the rapid fracture of lubricated frictional interfaces}
\affiliation{The Racah Institute of Physics, The Hebrew University of Jerusalem, Jerusalem, Israel}

\author{E.~Bayart} \affiliation{The Racah Institute of Physics, The Hebrew University of Jerusalem, Jerusalem, Israel}
\author{I.~Svetlizky} \affiliation{The Racah Institute of Physics, The Hebrew University of Jerusalem, Jerusalem, Israel}
\author{J.~Fineberg} \affiliation{The Racah Institute of Physics, The Hebrew University of Jerusalem, Jerusalem, Israel}
\date{\today}

\begin{abstract}
\noindent We study the onset of friction for rough contacting blocks whose interface is coated with a thin lubrication layer. High speed measurements of the real contact area and stress fields near the interface reveal that propagating shear cracks mediate lubricated frictional motion. While lubricants reduce interface resistances, surprisingly, they significantly increase energy dissipated, $\Gamma$, during rupture. Moreover, lubricant viscosity affects the onset of friction but has {\em no} effect on $\Gamma$. Fracture mechanics provide a new way to view the otherwise hidden complex dynamics of the lubrication layer. 
\end{abstract}

\pacs{46.55.+d, 46.50.+a, 62.20.Qp, 81.40.Pq}
\maketitle

Lubrication of solid surfaces is generally used to reduce frictional resistance to sliding motion and to prevent material wear \cite{Bowden2001}. Effects of fluids on the frictional properties of an interface are of particular significance in geophysics, since tectonic faults are generally lubricated by interstitial water or melted rocks \cite{Rice2006,Noda2009,Brantut2011,Viesca2015}. Along spatially extended multi-contact interfaces, which are considered here, much fundamental understanding of the collective mechanisms responsible for the reduction of friction due to lubrication is still lacking \cite{Rigney2010,Pastewka2011,Gosvami2015}. While the sliding dynamics of lubricated systems is an active field of research \cite{Persson2000,Campem2012,Ernesto2015}, the mechanisms mediating their transition from stick to slip remain largely unexplored. At the microscopic level, stick-slip mechanisms have been discussed for decades \cite{Thompson1990,Rosenhek2015}. Within single contacts, confined lubrication layers, typically at nanometric sizes, exhibit enhanced strength \cite{Granick1991,Klein1995,Bhushan1995}.

Along spatially extended rough interfaces, the real contact area is defined by a large ensemble of single contacts (asperities) that couple contacting elastic blocks.  The real contact area, $A$, is generally orders of magnitude smaller than the apparent one \cite{Bowden2001,Greenwood1966,Dieterich1996}. Here we consider rough surfaces in the boundary lubrication regime, where the contacting surfaces are covered by a thin lubricant layer \cite{Hamrock2004}. The discrete asperities in this regime still bear the entire normal load; they are not entirely immersed in the fluid layer as in the full lubrication regime.  The mixed lubrication regime is an intermediate region, where the normal load is partially borne by solid contacts and partially by the liquid layer.

In dry friction, the onset of motion is mediated by rupture fronts propagating along the frictional interface \cite{Rubinstein2004,Xia2004}. These fronts are true singular shear cracks; the strain fields during their propagation are well-described by Linear Elastic Fracture Mechanics (LEFM) \cite{Svetlizky2014}. Frictional rupture arrest is also governed by the same framework \cite{Kammer2015, Bayart2016}. 

Here, we examine the mechanisms coming into play when motion initiates within {\em lubricated} interfaces, in the boundary lubrication regime. We first find that interface rupture still corresponds to the shear cracks described by LEFM. While reducing static friction by facilitating rupture nucleation, we will show that, surprisingly, lubricants make solid contacts effectively tougher, increasing the fracture energy of the interface (the dissipated energy per unit crack extension). Moreover, while the macroscopic frictional resistance of the interface depends on the lubricant viscosity, the fracture energy does not. We use this to demonstrate that nucleation and propagation of frictional ruptures are independent processes.

\begin{figure}[ht]
\includegraphics[scale=1]{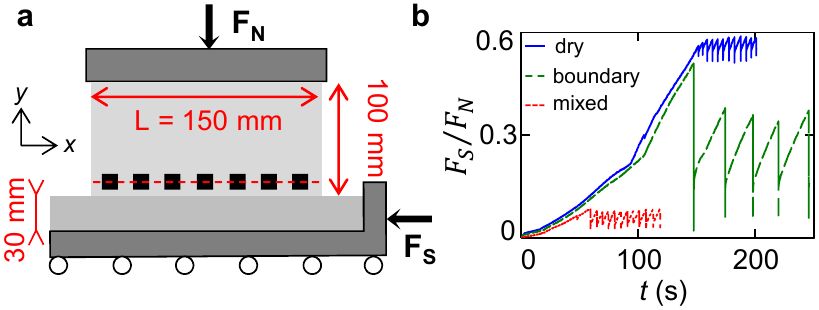}
\caption{\label{Fig1} Experimental setup and stick-slip behavior. (a) Normal  $F_N$ and shear  $F_S$ forces are applied to contacting PMMA blocks. Shear is applied uniformly via translation of a rigid stage. Strain gage rosettes measure the 3 components of the 2D-strain tensor at 14 locations along and 3.5 mm above the interface, while the real contact area is measured optically. (b) Loading curves, $F_S/F_N$ vs time, are plotted for typical experiments, with $F_N=4000\,\textrm{N}$: dry (solid blue line), boundary lubricated (dashed green line) and, for comparison, in the mixed lubricated regime (dotted red line). The lubricant used is a hydrocarbon oil (TKO-77).}
\end{figure} 
\begin{figure*}[ht]
	\includegraphics[scale=1]{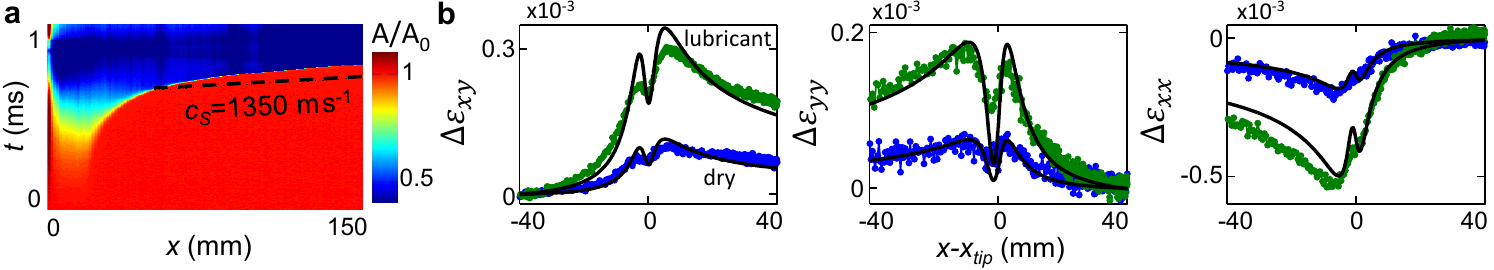}
	\caption{\label{Fig2} Singular interfacial shear cracks govern friction initiation. (a) The spatio-temporal evolution of the contact area $A(x,t)$ of a typical lubricated interface (hydrocarbon oil, TKO-77). Each line is a snapshot in time of $A(x,t)$, normalized by $A_0=A(x,0)$ immediately prior to the event. Here, a rupture accelerates to a propagation velocity $c_f=0.92\,c_R$. The rupture tip, $x_{tip}(t)$  are the locations where $A(x,t)$ drops sharply. (b) Variation of the strain field $\Delta\varepsilon_{ij}(x-x_{tip})$ with the distance from rupture tips, $x_{tip}$, for ruptures propagating along dry (blue line) and lubricated (green line, rupture presented in (a)) interfaces. In both, the applied normal stress was $\left\langle\sigma_{yy}\right\rangle=7\pm0.5$ MPa and strains were measured at $x=77\,\textrm{mm}$,  where $c_f\sim0.3\,c_R$. Black solid lines are fits to the LEFM solution for $y=3.5$ mm (Eq.~\ref{Eq1}). The only fitting parameter is the fracture energy; $\Gamma_{dry}= 2.6\pm0.3\,\textrm{J}\,\textrm{m}^{-2}$ for the dry and $\Gamma_{lub}=23\pm3\,\textrm{J}\,\textrm{m}^{-2}$ for the lubricated interfaces.}
\end{figure*}

We describe experiments where two blocks of poly(methylmethacrylate) (PMMA) are first pressed together with normal forces, $F_N$, of $2500<F_N<7000\,\textrm{N}$. Shear forces, $F_S$, are then applied uniformly, as the bottom block is translated via a rigid stage, until stick-slip motion initiates (Fig.~\ref{Fig1}). A detailed description of the setup is given in \cite{Svetlizky2014}. PMMA has a rate-dependent Young's modulus $ 3<E<5.6\,\textrm{GPa}$ and Poisson ratio $\nu_p=\!0.33$. PMMA's Rayleigh wave speed is $c_R\!=\!1255\:\textrm{m}\,\textrm{s}^{-1}$ for plane strain conditions. Top and bottom blocks have respective $x\times y\times z$ dimensions $150 \times 100\times 5.5\,\textrm{mm}$ and $200 \times 30\times 30\,\textrm{mm}$. The contacting flat surfaces of the top and bottom blocks were (top) optically smooth and (bottom) with a surface roughness of $0.5\,\mu\textrm{m}$ r.m.s. Experiments were all performed with the same two blocks, to negate any effects due to surface preparation or roughness. During each sliding event, an array of 14 strain gages recorded the 3 components of the 2D-strain tensor, $\varepsilon_{ij}$, 3.5 mm above the interface, each at $10^6$ samples/s. Corresponding stresses, $\sigma_{ij}$, are calculated from $\varepsilon_{ij}$ after accounting for the viscoelasticity of PMMA  (see \cite{Bayart2016}). In parallel, the real area of contact, $A(x,t)$, was measured at 1000 x 8 locations at 580000 frames/s, using an optical method based on total internal reflection (see \cite{Svetlizky2014}) where incident light only traverses the interface at contacts, and is otherwise reflected.

Experiments of lubricated friction were performed using silicone oils with kinematic viscosities, $\nu$= 5, 100 and $10^4$ $\textrm{mm}^2\,\textrm{s}^{-1}$ and a hydrocarbon oil (TKO-77, Kurt J. Lesker Company) of $\nu \sim 200$ $\textrm{mm}^2\,\textrm{s}^{-1}$. Lubricants were applied to either or both of the contacting surfaces and then wiped. Our results are not appreciably affected by the wiping procedure (number of wipes, application or not between experiments) or by the cleaning (soap, water and isopropanol). PMMA and the lubricants used are nearly index-matched: PMMA-1.49, TKO-77-1.48, and  silicone oils-1.42. Hence, under total internal reflection, incident light will be totally transmitted where gaps between asperities are filled with liquid. As light could be transmitted via capillary bridges across contacting surfaces, we only consider relative variations in light intensity. At the onset of motion, the observed contact area variations (see below), demonstrate that air, not lubricant, fills the gaps between contacts. This provides validation that the experiments take place in the boundary lubrication regime. 

When sheared, the lubricated system undergoes stick-slip motion (Fig.~\ref{Fig1}b). Drops of $F_S$ in the loading curves correspond to slip events with macroscopic relative displacement of the blocks. The lubricant layer affects the macroscopic frictional  resistance, reducing the static friction coefficient (i.e. the shear force threshold). The amplitudes of the force drops, however, are larger than for dry friction. In the boundary lubrication regime, this pattern is extremely robust, and is independent of the nature and quantity of the lubricant. For completeness, a typical loading curve in the mixed lubrication regime is included in Fig.~\ref{Fig1}b, where $F_S/F_N$ thresholds are further reduced. Motion in this regime is not addressed here.

As in dry friction, each sliding event in the boundary lubrication regime is preceded by propagating rupture fronts that break the solid contacts forming the interface, as shown in Fig.~\ref{Fig2}a. Macroscopic sliding only occurs when a front traverses the entire interface \cite{Rubinstein2007,Braun2009,Tromborg2011,Bayart2016}. For steady rupture fronts moving at a velocity $c_f$, $\varepsilon_{ij}(x,t)=\varepsilon_{ij}(x-c_f t)$. Using this and the optically identified location of the rupture tip, $ x_{tip}(t) $, we converted $\varepsilon_{ij}(x,t)$ to spatial measurements $\varepsilon_{ij}(x-x_{tip})$ \cite{Svetlizky2014}. As in the example of Fig.~\ref{Fig2}b (blue line), rupture fronts in dry friction are shear cracks whose stress field variations, $\Delta\sigma_{ij}(r,\theta)$, are quantitatively described by LEFM, with respect to the crack tip ($r=0$) \cite{Svetlizky2014}:
\begin{eqnarray}
\Delta\sigma_{ij}(r,\theta)=\frac{K_{II}(c_f)}{\sqrt{2\pi r}}\Sigma_{ij}^{II}(\theta,c_f),
\label{Eq1}
\end{eqnarray}
where $\Sigma_{ij}^{II}(\theta,c_f)$ is a universal angular function and the coefficient, $K_{II}(c_f)$, is called the stress intensity factor \cite{Freund1990}. $\Delta\sigma_{ij}$ expresses the stress changes between the initially applied and residual stresses along the frictional crack faces. $\Delta\sigma_{ij}$ are related to measured strain variations $\Delta\varepsilon_{ij}$ via the dynamic Young's modulus and Poisson ratio of PMMA.  LEFM relates $K_{II}$ to the fracture energy, $\Gamma$, the energy dissipated per unit crack advance; $K_{II} \propto f(c_f)\sqrt{\Gamma}$, where $f(c_f)$ is a known universal function \cite{Freund1990}.

\begin{figure}[ht]
	\includegraphics[scale=1]{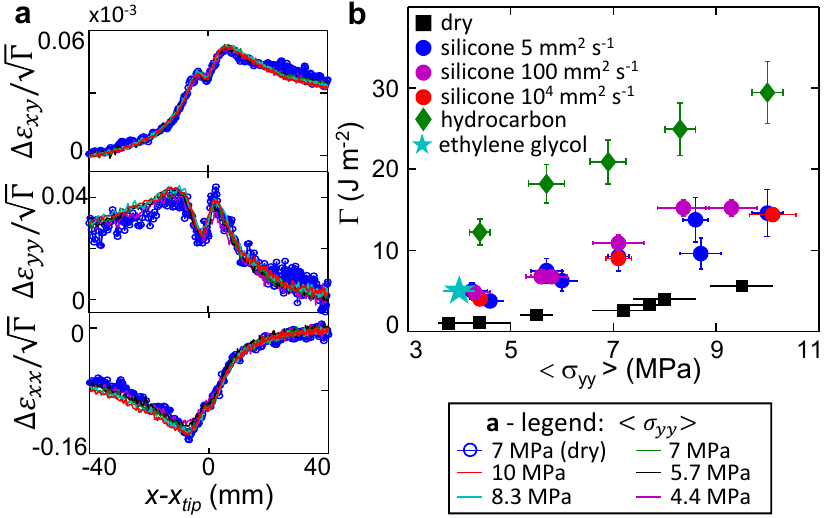}
	\caption{\label{Fig3} Dependence of the fracture energy with normal stress. (a) Comparison of $\Delta\varepsilon_{ij}(x-x_{tip})$ for dry and lubricated experiments, when normalized by $\sqrt{\Gamma}$, for different normal loads. Units are $(\textrm{Pa}\,\textrm{m})^{-1/2}$. Superimposed are the dry experiment in Fig.~\ref{Fig2}b and 5 lubricated (TKO-77) experiments where $c_f\sim 0.3\,c_R$ with $\left\langle\sigma_{yy}\right\rangle$  as in the legend, yielding $\Gamma_{dry}=2.6\,\textrm{J}\,\textrm{m}^{-2}$ and  $\Gamma_{TKO}=$ 12.3, 18.2, 23, 25 and 29.5 $\textrm{J}\,\textrm{m}^{-2}$. (b) $\Gamma$ is measured by fitting the strain field with the LEFM solution (as in (a)) for both dry and lubricated interfaces vs $F_N$. All $\Gamma$ vary linearly with $F_N$, $\Gamma$ is independent of the lubricant viscosity while highly dependent on lubricant composition.}
\end{figure}

In Fig.~\ref{Fig2}b we compare measurements of $\Delta\varepsilon_{ij}(x-x_{tip})$ during rupture front propagation for dry and lubricated interfaces. Following Eq.~\ref{Eq1}, fitting the three strain components provides a dynamic measurement of $K_{II}$ \cite{Freund1990,Svetlizky2014} and, therefore, a measurement of $\Gamma$. As Fig.~\ref{Fig2}b explicitly shows, the agreement between measured $\Delta\varepsilon_{ij}(x-x_{tip})$ for the {\em lubricated} interface and the LEFM solution is excellent. Hence, ruptures propagating along a lubricated interface are shear cracks. Surprisingly, $\Gamma$, for the same applied normal load, is an order of magnitude {\em greater} for the lubricated interface, $\Gamma_{lub}$, than for the dry one, $\Gamma_{dry}$. For the examples presented in Fig.~\ref{Fig2}b, $\Gamma_{lub}=23\pm3\:\textrm{J}\,\textrm{m}^{-2}$ while $\Gamma_{dry}=2.6\pm0.3\:\textrm{J}\,\textrm{m}^{-2}$. In Fig.~\ref{Fig3}a we present $\Delta\varepsilon_{ij}(x-x_{tip})$ for dry and lubricated (hydrocarbon) interfaces, when rescaled by $1/\sqrt{\Gamma}$. We find that the rescaled dry and lubricated strain fields are indeed {\em identical}.

What determines $\Gamma$? In dry friction, when contacts are plastically deformed, $\Gamma$ grows linearly with the normal load \cite{Bowden2001,Bayart2016}. Extracting $\Gamma$ from the rescaling procedure, Fig.~\ref{Fig3}b shows that $\Gamma$ indeed remains proportional to the average normal stress, $\left\langle\sigma_{yy}\right\rangle$, in the boundary lubrication regime. Moreover, the value of $\Gamma$ is {\em unaffected} by the lubricant viscosity; $\Gamma$ is {\em constant} for viscosity variations of $5<\nu<10^4\,\textrm{mm}^{2}\,\textrm{s}^{-1}$ in silicon oils. We do, however, find that $\Gamma$ strongly depends on the lubricant composition; TKO-77 has values of $\Gamma$ about 3 times larger than in all of the silicon oils used. For a given $c_f$, increased values of $\Gamma$ induce increased shear stress drops during rupture propagation. The increased shear force drops in loading curves (e.g. Fig.~\ref{Fig1}b) are partially caused by this large stress drop, with the remainder due to motion after the rupture passage. 

\begin{figure}
	\includegraphics[scale=1]{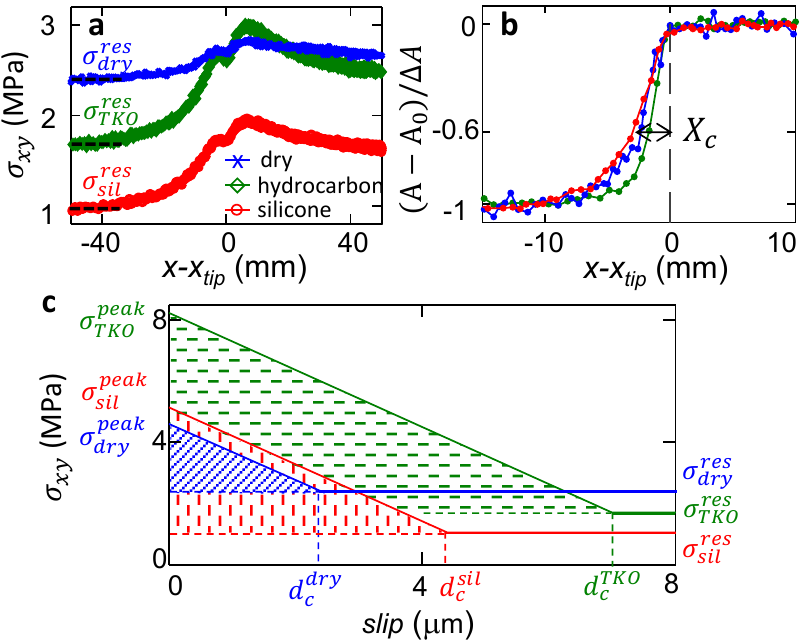}
	\caption{\label{Fig4} Measurements of $\sigma_{xy}^{res}$, $X_c$, $\sigma_{xy}^{peak}$ and $d_c$. (a) Shear stress as a function of the distance from $x_{tip}$ for ruptures propagating ($c_f\sim 0.3c_R$) along a dry (blue crosses) and lubricated interfaces with hydrocarbon oil (green diamonds) and silicone oil with $\nu=10^4\,\textrm{mm}^2\,\textrm{s}^{-1}$ (red circles). $\Gamma=$2.6, 9 and 23 $\textrm{J}\,\textrm{m}^{-2}$ for respectively dry, silicone and hydrocarbon oils for $\left\langle\sigma_{yy}\right\rangle=$ 7 MPa. The blue and green plots are measurements presented in Fig.~\ref{Fig2}b, where stain variations are, instead, presented in terms of the absolute stress values. (b) Reduction of the contact area $A-A_0$, normalized by the total drop in $A$, $\Delta A = A_0-A_{res}$, as a function of the distance from $x_{tip}$ for the three experiments in (a). The dissipative zone size $X_c$ is defined as the length scale where a 60\% drop of $\Delta A$ occurs. (c) Shear stress vs slip distance where $\sigma_{xy}^{peak}$ and $d_c$ are estimated within the linear slip-weakening model \cite{Palmer1973}. Respectively for dry and lubricated with silicone and hydrocarbon oils interfaces, residual stresses, defined in (a), are 2.4, 1 and 1.7 MPa, peak stresses $\sigma_{xy}^{peak}$, using Eq.~\ref{Eq2}, are 4.6, 5.1 and 8.2 MPa and $d_c$ are 2.4, 4.4 and 7 $\mu$m. Integration over the blue (green, red) hatched areas provides the dry (lubricated) fracture energy.}
\end{figure}

Why does the lubricant {\em increase} $\Gamma$? We consider the simplest (linear slip-weakening) description of the dissipative zone near a rupture tip \cite{Palmer1973}. In this model, rupture occurs when the shear stress on the interface reaches a maximal value, $\sigma_{xy}^{peak}$. Slip is then initiated and $\sigma_{xy}$ is reduced to the residual value $\sigma_{xy}^{res}$ over a slip distance $d_c$. While simple, this model contains the main features of the regularized dissipative zone, and $d_c$ provides an accurate estimate of the sliding distance, typically the asperity size. The fracture energy is expressed as $\Gamma=\frac{1}{2}(\sigma_{xy}^{peak}-\sigma_{xy}^{res})d_c$. More sliding occurs after rupture passage, dissipating more energy. Therefore, the energy dissipated by the rupture is only part of the total energy dissipated during a slip event.
An increase of $\Gamma$ can be induced by increased values of either $\sigma_{xy}^{peak}$ or $d_c$, or a decrease of $\sigma_{xy}^{res}$. 

As Fig.~\ref{Fig4}a shows, $\sigma_{xy}^{res}$ is indeed strongly reduced by the lubricant. The magnitude of the reduction relative to the dry interface depends on the nature of the lubricant. It is greater for silicone oil than for TKO-77. We can not measure $\sigma_{xy}^{peak}$ directly, as our strain gages are located above the interface \cite{Svetlizky2014}. The linear slip weakening model provides us with a way to accurately estimate $\sigma_{xy}^{peak}$ by measuring the size of the dissipative zone $X_c$, the distance behind the crack tip over which contacts are being broken \cite{Palmer1973}:
\begin{eqnarray}
\sigma_{xy}^{peak}=\sigma_{xy}^{res}+\sqrt{\frac{9\pi}{32}\frac{\Gamma E}{(1-\nu_p^2)X_c}}
\label{Eq2}
\end{eqnarray}
$X_c$ is the scale over which $A(x)$ drops from its initial to residual value. In Fig.~\ref{Fig4}b we compare $X_c$ for dry and lubricated experiments (see \cite{Svetlizky2014} for details). We find that $X_c$ is not significantly affected by the lubricant layer; its value (for $c_f\sim0.3c_R$) is approximately 3 mm. Inserting this value in Eq.~\ref{Eq2}, we see that $\sigma_{xy}^{peak}$ is \textit{not} reduced by the lubricant and is even significantly increased when TKO-77 is used (Fig.~\ref{Fig4}c). The dynamic measurements of the stress drop, $\sigma_{xy}^{peak}-\sigma_{xy}^{res}$ coupled with $\Gamma$ yield a quantitative estimate of $d_c$. Measurements of $\sigma_{xy}^{peak}$, $\sigma_{xy}^{res}$, $\Gamma$ and $d_c$ are presented Fig.~\ref{Fig4}c. These measurements indicate that increases in $\Gamma$ are therefore explained by increased stress drops, coupled to larger slip distances.

\begin{figure}
	\includegraphics[scale=1]{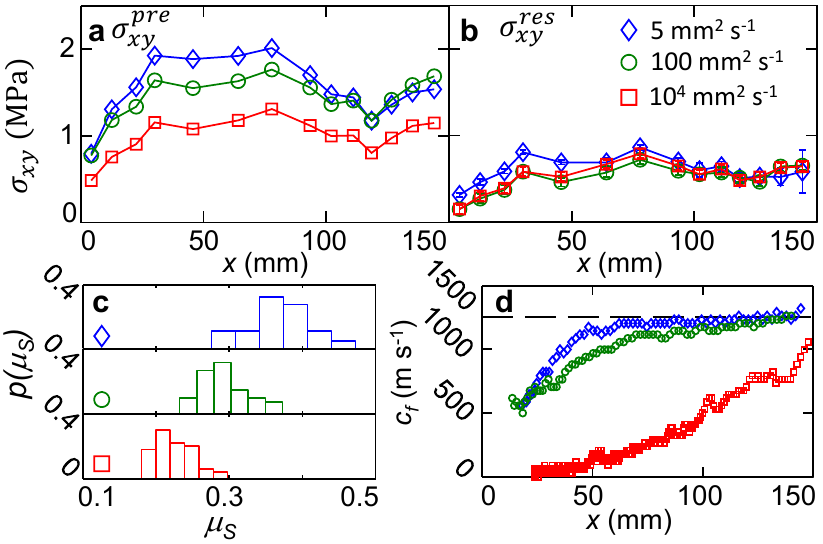}
	\caption{\label{Fig5} Dependence of the macroscopic frictional resistance with $\nu$. Profiles of (a)  the initial $\sigma_{xy}^{pre}$  and (b) residual stresses $\sigma_{xy}^{res}$ for interfaces lubricated with silicone oils having $\nu =5$ (diamonds), 100 (circles) and $10^4$ (squares) $\textrm{mm}^2\,\textrm{s}^{-1}$. Normal stress distributions are identical for the three experiments and $\Gamma=6.7\pm0.3\,\textrm{J}\,\textrm{m}^{-2}$. $\sigma_{xy}^{res}$ does not depend on the viscosity of the silicone oil. (c) Histograms of the static friction coefficient $\mu_S$ of sliding events for lubricated interfaces with silicone oils of different viscosity. Around 50 events are considered for each $\nu$. Higher $\nu$ yield lower frictional resistance. (d) 3 examples of $c_f(x)$ for the lubricants in (a) and (b). The larger $\nu$, the slower the front. Dashed line denotes $c_R$. Symbols and colors in (a-d) as in (b).}
\end{figure}

We have seen that, in the boundary lubrication regime, while the contacts become tougher, requiring a larger amount of energy ($\Gamma$) to break (Fig.~\ref{Fig3}), the macroscopic frictional resistance is actually reduced (Fig.~\ref{Fig1}b). These intriguing results are not contradictory; we show here that rupture nucleation and dissipation are independent processes. 

Rupture nucleation determines the initial stress levels and therefore, an interface's ``static" frictional strength. Hence, nucleation is the key in understanding initial interfacial strength, although processes determining how and at what stress levels nucleation takes place remain enigmatic \cite{Yang2008,Latour2013}. Fig.~\ref{Fig5} demonstrates that the lubricant viscosity directly affects the initially imposed stress, $\sigma_{xy}^{pre}$, needed to nucleate the rupture. Using silicone oils of different viscosities, in experiments performed with the same normal stress profile, Fig.~\ref{Fig5} reveals that the higher $\nu$, the lower $\sigma_{xy}^{pre}$. On the other hand, $\sigma_{xy}^{res}$ does {\em not} depend on $\nu$. $\sigma_{xy}^{pre}$ determines the static friction coefficient $\mu_s=F_S/F_N$. Hence, as Fig.~\ref{Fig5}c shows, $\mu_s$ is significantly dependent on $\nu$, as suggested by earlier studies \cite{Hardy1936,Bowden2001}. 

As $\Gamma$ and $\sigma_{xy}^{res}$ are $\nu-$independent (Figs.~\ref{Fig3} and \ref{Fig5}b), LEFM predicts that the only effect of $\sigma_{xy}^{pre}$ should be on the rupture dynamics. A larger $\sigma_{xy}^{pre}$ yields faster rupture fronts \cite{Freund1990} as verified in both dry friction \cite{Ben-David2010s} and in ice-quakes \cite{Walter2015}. This is born out by the examples shown in Fig.~\ref{Fig5}d; for the same $\sigma_{yy}$ profile, the higher $\nu$, the lower $\sigma_{xy}^{pre}$, and the slower the rupture front. Our results imply that the reduction of $\mu_s$  is purely the result of rupture front nucleation at a reduced threshold, due to higher $\nu$.
Lower initial stresses do not prevent interfacial rupture propagation as long as the elastic energy stored by block deformation is sufficiently above the energy dissipated during the rupture process (Fig.~\ref{Fig4}c).

We have shown that, in the boundary lubrication regime, interfacial resistance is reduced due to facilitated nucleation of the rupture front (Fig.~\ref{Fig5}), while contacts become tougher (Fig.~\ref{Fig3}). The increase of $\Gamma$ is explained by a strengthening of the contacts (increased $\sigma_{xy}^{peak}$) coupled to reduced $\sigma_{xy}^{res}$, with resultant increases in slip distances $d_c$. The reduced $\sigma_{xy}^{res}$ and larger $d_c$ may result from the fact  that, while in motion, the lubricant layer facilitates slip. 
The increased value of $ \Gamma $ and $ \sigma^{peak}_{xy} $, however, are both new and intriguing observations. In our experiments, on rough multi contact interfaces, pressures at a single contact reach the yield stress of PMMA ($\sim$500 MPa) \cite{Dieterich1996}, suggesting that nanometric lubricant layers could well be trapped between asperities. At these extreme conditions the contribution of capillary bridges is negligible to both the frictional resistance and $\Gamma $ (see Supp. Mat.).
At a microscopic level, the physics of lubricated single contact interfaces are both interesting and puzzling. Experiments reveal that fluid lubrication layers confined to nanometric scales can transition to solids \cite{Granick1991,Klein1995,Bhushan1995}. Other recent experiments on similar systems suggest that the nature of fluid layers does not change, but high system stiffness results from coupling of the fluid to elastic deformation of the surrounding medium \cite{Villey2013}. While our rough, multi-contact system is far from these ideal cases, it is interesting that lubricant strengthening indeed takes place. Obtaining a fundamental understanding of the dynamics of the lubrication layer and its associated dissipative properties in such a disordered system is an important and interesting challenge. 

We acknowledge support from the James S. McDonnell Fund Grant 220020221, the European Research Council (Grant No. 267256) and the Israel Science Foundation (Grants 76/11 and 1523/15). E.B. acknowledges support from the Lady Davis Trust. We thank G. Cohen for comments.

\bibliography{references}

\end{document}